\documentclass[superscriptaddress,showpacs,aps,prd,notitlepage]{revtex4-1}
\usepackage{epsfig}
\usepackage{amsmath,amssymb}
\usepackage{color}
\usepackage{booktabs}
\usepackage{chngpage}
\usepackage{float}
\usepackage{setspace}
\RequirePackage{graphicx,subfigure}
\begin{document}
%\preprint{APS/123-QED}
\title{Effective perihelion advance and  potentials in a conformastatic background with magnetic  field }

\author{Abra\~{a}o J. S. Capistrano }
\email{abraao.capistrano@unila.edu.br} \affiliation{Federal University of Latin-American Integration, 85867-970, P.o.b: 2123, Foz do Iguassu, PR, Foz do Igua\c{c}u-PR, Brazil \\
Casimiro Montenegro Filho Astronomy Center, Itaipu Technological Park, 85867-970, Foz do Iguassu, PR, Foz do Igua\c{c}u-PR, Brazil}

\author{Antonio C. Guti\'errez-Pi\~{n}eres}
\email[e-mail:]{acgutierrez@correo.nucleares.unam.mx}
\affiliation{Instituto de Ciencias Nucleares, Universidad Nacional Aut\'onoma de M\'exico,
 \\AP 70543,  M\'exico, DF 04510, M\'exico}
\affiliation{Facultad de Ciencias B\'asicas,\\
Universidad Tecnol\'ogica de Bol\'ivar, Cartagena 13001, Colombia}

\begin{abstract}
An Exact  solution of  the Einstein-Maxwell field equations for  a  conformastatic  metric  with  magnetized
sources is study. In this context, effective potential are studied in order to understand the dynamics of the magnetic field in galaxies. We derive the  equations of motion for  neutral and charged  particles in a spacetime background characterized  by  this  class of solutions. In this particular case, we investigate the main physical properties of equatorial circular orbits and related effective potentials. In addition, we obtain an effective analytic expression  for the  perihelion advance of test particles. Our theoretical predictions are compared with the observational data calibrated with the ephemerides of the planets of the Solar system and the Moon (EPM2011). We show that, in general, the magnetic punctual mass predicts values that are in better agreement with observations than the values predicted in Einstein gravity alone.
\end{abstract}

%\pacs{04.20.-q, 04.20.Jb,  04.40.-b,  04.40.Nr}

\maketitle
\section{Introduction}
Magnetics field are extensively studied in literature and its influence on the dynamics are on currently field of research, e.g, on the understanding the galactic jets and inner process of ``active'' galaxy core, neutron stars dynamics \cite{Bocquet}.  A interesting review can be found in \cite{Han,Han2,Krause,Beek1,Beek2} and/or movement of charged particles in spacetimes \cite{Hackman, Chakraborty,Nandi} or neutral particles in charged galactic halo \cite{Pugliese1,Pugliese2,Pugliese3}. The Einstein Maxwell equations have been revealed to be an important tool to deal with this problem and helping us on the understanding the dynamics of magnetic fields in galaxies. Important approaches are the relativistic models with disk like configurations and relativistic disk accretion models proposed in recent years, e.g, \cite{Voigt,Antonio1,Antonio2,Binney,Antonio3,Capistrano} and references therein. In a recent publication \cite{Antonio2}, we studied the behaviour of a test particle submitted to a magnetic field in a relativistic galaxy disk model and on how its influence may affect its dynamics. In this paper, we investigate effective potentials using Einstein-Maxwell equations motivated by the necessity to understand how the dynamics of a galaxy responds to flattening and how the magnetic field can be related to this process.

The present paper is divided in sections. In the second section, we study the basic framework of a conformastatic background and investigate some applications using the isothermal-sphere logarithm potential and Toomre-Kuzmin-like potential, which are compatible with axisymmetric systems. In the third section, we obtain an expression for the  perihelion advance of a charged test particle  in a generic  conformastatic spacetime  in the presence of  a magnetic field and perform a comparison between our  results, the results from Einstein gravity alone and the values observed for the secular perihelion precession of some inner planets and minor objects of the Solar System. In the conclusion section, we make the final considerations.

\section{The conformastatic background}
Considering  the  background of  a conformastatic gravitational source  in  presence of a  magnetic  field described  by  the  line  element  in standard cylindrical coordinates
 \cite{Antonio1},
   \begin{align}
    dS^2= -e^{2\phi}dt^2 + e^{-2\phi} (dr^2 + dz^2 + r^2d\varphi^2), \label{eq:metric}
   \end{align}
where the  metric  potential $\phi$ depend only on the  $r$  and  $z$. We use the term conformastatic in the sense of Synge \cite{Synge}. The  vacuum Einstein-Maxwell equations   in geometrized units such that $c = 8\pi G= \mu _{0} =
\epsilon _{0} =  1$, are  given  by
  \begin{subequations}\begin{align}
   G_{\alpha\beta}& = E_{\alpha\beta},\\
   F^{\alpha\beta}_{\;\;\;\; ; \beta}&= 0,
   \end{align}\label{eq:E-M}\end{subequations}
 where  $F_{\alpha\beta} =  A_{\beta,\alpha} -  A_{\alpha, \beta}$  and    $E_{\alpha\beta}$ is the electromagnetic energy-momentum tensor
       \begin{align*} E_{\alpha\beta} = \frac{1}{4 \pi} \left\{ F_{\alpha\gamma}F_{\beta}^{ \; \gamma}
                                                        -  \frac{1}{4}g_{\alpha\beta}F_{\gamma\delta}F^{\gamma\delta} \right\}. \label{eq:tab}
          \end{align*}
The Greek indices run from 1 to 4.

With the electromagnetic  potential $A_{\alpha}= (0,0,0, A_{\varphi}(r,z))$ and  the  line element in Eq.(\ref{eq:metric}) the  Einstein-Maxwell equations in Eq.(\ref{eq:E-M}) are equivalent to the system of equations
   \begin{subequations}\begin{align}
    \nabla\cdot \left( r^{-2} e^{2\phi} \nabla A_{\varphi} \right) =0 , \\
    \nabla^2 \phi - \nabla \phi \cdot\nabla \phi =0, \\
     \phi_r^2 - \frac{1}{2r^2} e^{2\phi} A_{\varphi,z}^2 =0,\\
     \phi_z^2 - \frac{1}{2r^2} e^{2\phi} A_{\varphi,r}^2 =0,\\
     \phi_r \phi_z + \frac{1}{2r^2}e^{2\phi}A_{\varphi,r}A_{\varphi,z}=0.
   \end{align}\label{eq:E-Mexplicitly}\end{subequations}
By  using  the  procedure  to  obtain solutions of  the  Einstein-Maxwell equations presented  in \cite{Antonio1}, a suitable solutions of  the  system
in Eq.(\ref{eq:E-Mexplicitly}) can be  displayed as
     \begin{subequations}\begin{align}
                       e^{\phi}&= \frac{1}{1-U},\\
             A_{\varphi,r}&=  \sqrt{2}r U_{,z},\\
             A_{\varphi,z}&= -\sqrt{2}r U_{,r},
                         \end{align}\label{eq:solutions}\end{subequations}
where $U(r,z)$ is  a  solution of  the Laplace's equation.

\subsection{Motion of  test charged  particles}
The  motion  of  a  test particle of  charge $q$ and  mass $m$  moving in  a magnetized  background  is  described  by  the  Lagrangian
  \begin{align}
   {\cal L} = \frac{1}{2} m   u_{\alpha} u^{\alpha} + q A_{\alpha} u^{\alpha},\label{eq:Lagrangian}
  \end{align}
 where $u^{\mu}= {d x^\mu}/{ds}$,  being  $s$ and  arbitrary  parameter. The  corresponding Hamiltonian   of  the  particle  is
   \begin{align}
   {\cal H} = \frac{1}{2m} (p^{\mu}  - q A^{\mu}) (p_{\mu}  - q A_{\mu}),\label{eq:Hamiltonian}
  \end{align}
where  the canonical momentum is  given  by $p_{\mu} = m u_{\mu}  + q A_{\mu}$. The motion  equations  are  given  by
                  \begin{subequations}
                  \begin{align}
                    u^{\mu}&= \frac{\partial  {\cal H} g^{\mu\nu} } {\partial p^{\nu} },\\
                    \frac{d p^{\mu}}{ds}&= - \frac{\partial  {\cal H} g^{\mu\nu} } {\partial p^{\nu} },
                  \end{align}\label{eq:Hamiltonequations}
                  \end{subequations}
where ${\cal H}_c \equiv p_{\mu} p^{\mu}/(2m)$. Accordingly,  by  introducing Eq.(\ref{eq:Hamiltonian}) into Eq.(\ref{eq:Hamiltonequations}) we obtain
                \begin{subequations}\begin{align}
                    \frac{d p^{t}}{ds}&=0,\label{eq:motionequations1}\\
                    \frac{d p^{\varphi}}{ds}&=0,\label{eq:motionequations2}\\
                    \frac{d p^{r}}{ds}&= - \frac{p^{\mu}p_{\mu}}{2m}  \frac{\partial g^{rr}}{\partial r},\label{eq:motionequations3}\\
                     \frac{d p^{z}}{ds}&= - \frac{p^{\mu}p_{\mu}}{2m}  \frac{\partial g^{zz}}{\partial z} \label{eq:motionequations4}.
                   \end{align}\label{eq:motionequations}\end{subequations}
From Eq.(\ref{eq:Hamiltonian}) and the  normalization  condition $u^{\mu}u_{\mu} = - \varepsilon$  (with $\varepsilon =(1,0,-1)$ for space-like, null and time-like curves) we have the  condition
                   \begin{align}
                    {\cal H}= -\frac{1}{2} m \varepsilon. \label{eq:HamiltonianEXPLIC}
                   \end{align}
 On the  another  hand,  from Eq.(\ref{eq:motionequations1}) and Eq.(\ref{eq:motionequations2})  we  have,
                       \begin{align}
                       p^{t} = \mbox{constant} \equiv - E,
                        \end{align}
 and also
                      \begin{align}
                       p^{\varphi} = \mbox{constant} \equiv L,
                        \end{align}
respectively. Whereas,  from Eq.(\ref{eq:motionequations3})  and Eq.(\ref{eq:motionequations4}) we  obtain
                         \begin{subequations}\begin{align}
                         \ddot{r} &= W   \frac{\partial g^{rr}}{\partial  r},\\
                         \ddot{z} &= W   \frac{\partial g^{rr}}{\partial  r},
                         \end{align}\end{subequations}
where    
\begin{align}
         W \equiv  \frac{1}{2} \Big(  \varepsilon + \frac{q^2}{m^2} A_{\varphi}A^{\varphi} - \frac{2qL}{m^2} A_{\varphi} \Big).\label{eq:W}
\end{align}

We  can write  the last  system  in the  form
                         \begin{subequations}\begin{align}
                         \ddot{r} &=  - \frac{\partial \Phi_{_{ eff}} }{\partial  r},\\
                         \ddot{z} &=  - \frac{\partial \Phi_{_{ eff}} }{\partial  r},
                         \end{align}\end{subequations}
where
                   \begin{align*}
                    d \Phi_{_{ eff}} = W   \frac{\partial g^{rr}}{\partial  r} dr + W   \frac{\partial g^{rr}}{\partial  z} dz.
                     \end{align*}
$\Phi_{_{eff}} $  is  called  the  ``effective  potential''  (See  equations (3.68)  Pg. 160  in  \cite{Binney}). In  terms  of the
solution in Eq.(\ref{eq:solutions}), one obtains
                         \begin{subequations}\begin{align}
                          \ddot{r} &= -  \frac{\partial \Phi_{_{eff}}(U)}{\partial r},\\
                           \ddot{z} &=  -  \frac{\partial \Phi_{_{eff}}(U)}{\partial z}
                          \end{align}\end{subequations}
where
                    \begin{eqnarray}\label{eq:effpotencial}
                    d \Phi_{_{eff}}(U)  = - \frac{ h(r,z)}{(1 - U)^3} dU,\;\;\;\;\;\;\;\;\;\;\;\;\;\;\;\;\;\;\;\;\;\\
                    h(r,z) \equiv \varepsilon +  \frac{ 2}{ m^2r^2 (1 - U)^2} \left(\frac{\partial  }{\partial  z} \int_0^r{ U r dr}\right)^2
                                                               - \frac{ 2 \sqrt{2} qL}{ m^2} \frac{\partial }{\partial  z} \int_0^r{U r dr}.
                      \end{eqnarray}
Thus  the three-dimensional  motion  of the  particle in an axis-symmetric potential can be  reduced  to the two-dimensional motion of the particle in a  ``Newtonian potential'' $U(r,z)$.

\subsection{Circular Motion in the plane $z=0$}
To  study  the  circular motion  of  the  test  charged  particle  we  start  with the  conditions
         \begin{align}
            \dot{r}=0, \quad \frac{ \partial \Phi_{_{eff} }}{\partial r} =0.\label{eq:firstconditions*}
          \end{align}
Then,  from the  first of  these equations, Eqs.(\ref{eq:Hamiltonian}) and (\ref{eq:HamiltonianEXPLIC}), we  have the energy of  the particle
                   \begin{align}
                     E^2= - g^{tt} \left(\varepsilon m^2 + g_{\varphi\varphi} (L - q A_{\varphi}  g^{\varphi\varphi})^2    \right).\label{eq:energy*}
                   \end{align}
From the  second  condition in Eq.(\ref{eq:firstconditions*})  we  have
         \begin{align}
            \ddot{r} = W   \frac{\partial g^{rr}}{\partial  r} =  - \frac{ \partial \Phi_{_{eff} }}{\partial r}=0.\label{eq:firstconditions}.
          \end{align}
Notice  that if  $W=0$ ,  from Eqs.(\ref{eq:energy*}) and (\ref{eq:W}),  we  obtain
         \begin{align}
          E^2=\frac{g_{\varphi\varphi}}{g_{tt}}L^2.
         \end{align}
Thus, by  introducing the corresponding metric coefficients of the  line element in Eq.(\ref{eq:metric}), such as
         \begin{align}
          E^2=-r^2 e^{-4\phi}L^2,
         \end{align}
which lacks of physical meaning. Hence, the  condition  $\frac{ \partial \Phi_{_{eff} }}{\partial r} =0$
is  equivalent to
      \begin{align}
      \frac{ \partial g^{rr} }{\partial r} =0,\quad  W \neq 0.
     \end{align}

The  minimum  radius for  stable circular  orbit occurs  in the  inflection  points  of the  effective  potential.  Thus  we  must  solve the  equation
            \begin{align}
           \ddot r= \frac{\partial ^2 \Phi_{_{eff}} }{\partial r^2} =0.
              \end{align}
              or,  equivalently, to solve the equation
       \begin{align}
            \frac{\partial ^2 g^{rr} }{\partial r^2} =0.
           \end{align}
On the other  hand,  by  calculating the  derivative  respect to  $z$ in  both  sides of Eq.(\ref{eq:energy*}) we  obtain for  the  angular moment
         \begin{align}
          L= q A_{\varphi} g^{\varphi\varphi} + \frac{l}{ (g^{tt} g_{\varphi\varphi})_{,z}},
         \end{align}
where
         \begin{align*}
          l \equiv q A_{\varphi} g^{tt}g_{\varphi\varphi} g^{\varphi\varphi}_{\;\;\;,z} \pm
          \sqrt{ \left(q A_{\varphi} g^{tt}g_{\varphi\varphi} g^{\varphi\varphi}_{\;\;\;,z} \right)^2
               -  \varepsilon m^2 g^{tt}_{\;\; ,z} (g^{tt} g_{\varphi\varphi})_{,z}},
         \end{align*}
and  we  have  used  the  Einstein-Maxwell  equation $\phi_r^2 = \frac{1}{2r^2} e^{2\phi} A_{\varphi,z}^2$. By  substituting  this  value for  $L$ in  (\ref{eq:energy*}) we  obtain the  energy  of  the  particle
           \begin{align}
                     E^2= - g^{tt} \left(\varepsilon m^2 + g_{\varphi\varphi} \frac{l^2}{ (g^{tt} g_{\varphi\varphi})_{,z}^2}    \right).\label{eq:energy}
                   \end{align}
Since  the  Lagrangian in Eq.(\ref{eq:Lagrangian}) does not  depend  explicitly  on the  variables $t$ and $\varphi$, and one can obtain the  following  two conserved  quantities
 \begin{equation}
   p_t= -mc e^{2\phi}\dot{t} \equiv -\frac{E}{c},
   \label{eq:energy}
\end{equation}
and also
\begin{equation}
   p_{\varphi}= m r^2 e^{-2\phi}\dot{\varphi} + \frac{q}c{}A_{\varphi} \equiv L,
   \label{eq:angularmomentum}
\end{equation}
where $E$ and $L$ are, respectively,  the  energy and the  angular  momentum  of  the particle  as  measured by  an observer at rest at infinity.
Furthermore, the momentum $p_{\alpha}$ of  the  particle  can be  normalized  so that $ g_{\alpha\beta} \dot x^{\alpha} \dot x^{\beta} = - \Sigma$. Accordingly, for the  metric in Eq.(\ref{eq:metric}) we have
\begin{equation}
 -{e^{2\phi} \dot{t}^2} + e^{-2\phi}(\dot{r}^2  + \dot{z}^2 + r^2 \dot{\varphi}^2  ) =- \Sigma,
 \label{eq:circuprecess}
\end{equation}
where, with $ c=1$, the notation $\Sigma = 1, 0, -1$  denotes space-like, null and time-like curves, respectively.

As an application of Eq.(\ref{eq:effpotencial}), we use an axial bi-dimensional isothermal potential, which has the form
$$U(r)= 1-v_0^2 \ln (r^2+z^2)\;,$$
and straightforwardly, we get the expression
$$d\Phi_{eff}= \frac{4}{v_0^4}\{\epsilon+ a\left(\frac{z}{r}\right)^2 + b z \ln(z^2+r^2)\} \frac{rdr+zdz}{(z^2+r^2)\ln(z^2+r^2)^3}       \;.$$
Hence, integrating the former expression, it is necessary to obtain a convergence of the integral away from origin, we use a Laurent expansion
$\sum^{\infty}_{k=1} = \frac{1}{k^2}\sim \frac{\pi^2}{6}$. Finally, after a long algebra, we can write the form of the effective potential felt by charged particle with mass moving with velocity $v_0$ and total angular momentum $L$
$$\Phi_{eff}(r,z)= \frac{\epsilon}{4v_0^4} \left[\frac{\ln^2(z^2+r^2) - 4 \ln^2 z}{\ln^2 z  \ln^2(z^2+r^2)} \right] + \frac{bz}{v_0^4} \left[\frac{\ln(z^2+r^2) - 2\ln z}{\ln z  \ln(z^2+r^2)} \right] $$
$$-\frac{a z^2}{v_0^4 r^2 \ln^2(z^2+r^2)} +\frac{a z^2 \pi^2}{24 v_0^4 \ln^2(z^2+r^2)} $$
where we denote $a= \frac{2v_0^2}{m^2}$ and $b= \frac{2\sqrt{2}}{m^2}qLv_0^2$.
\begin{figure}[!htb]
\minipage{0.32\textwidth}
  \includegraphics[width=\linewidth]{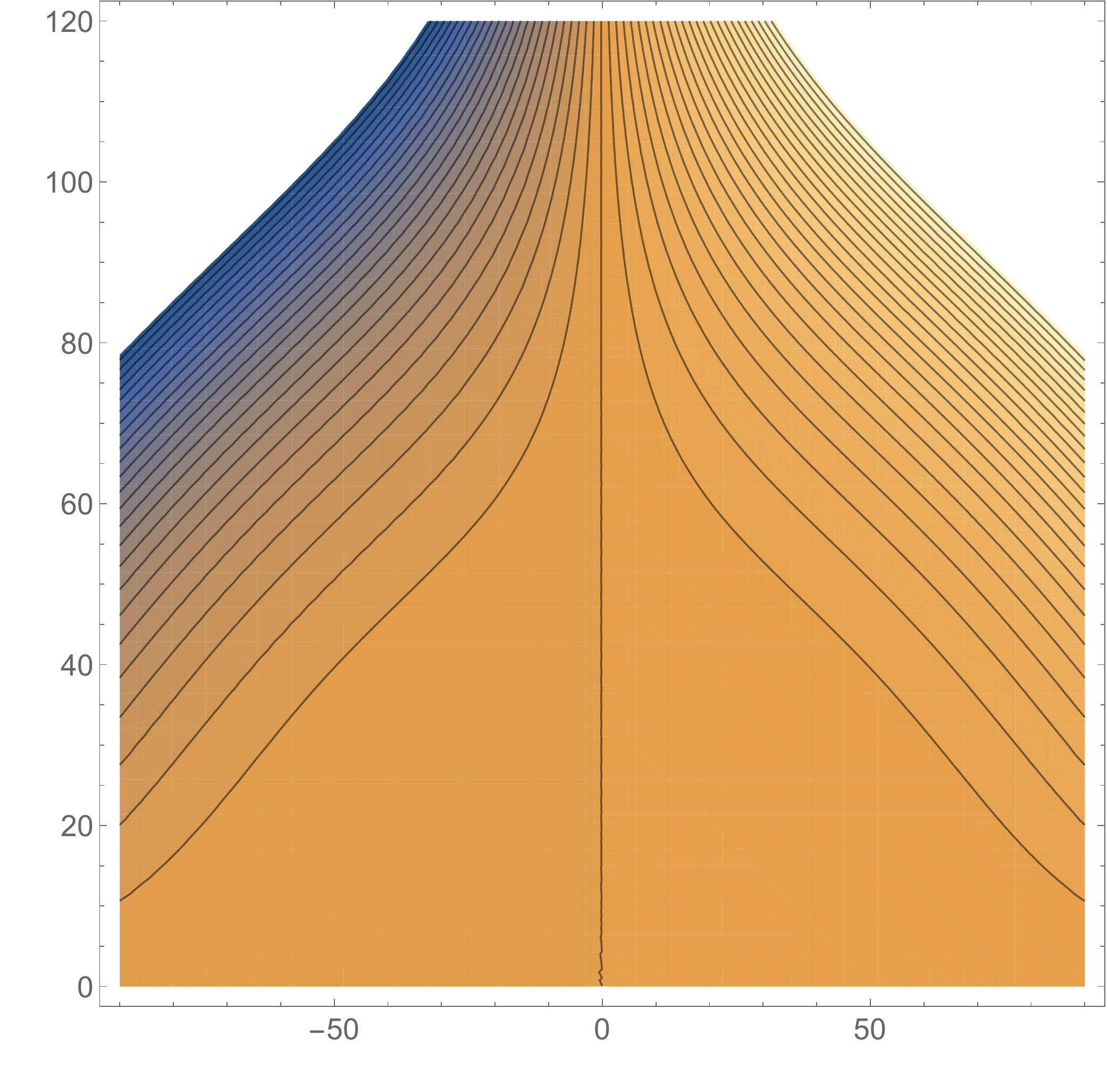}
  \endminipage\hfill
\minipage{0.32\textwidth}
  \includegraphics[width=\linewidth]{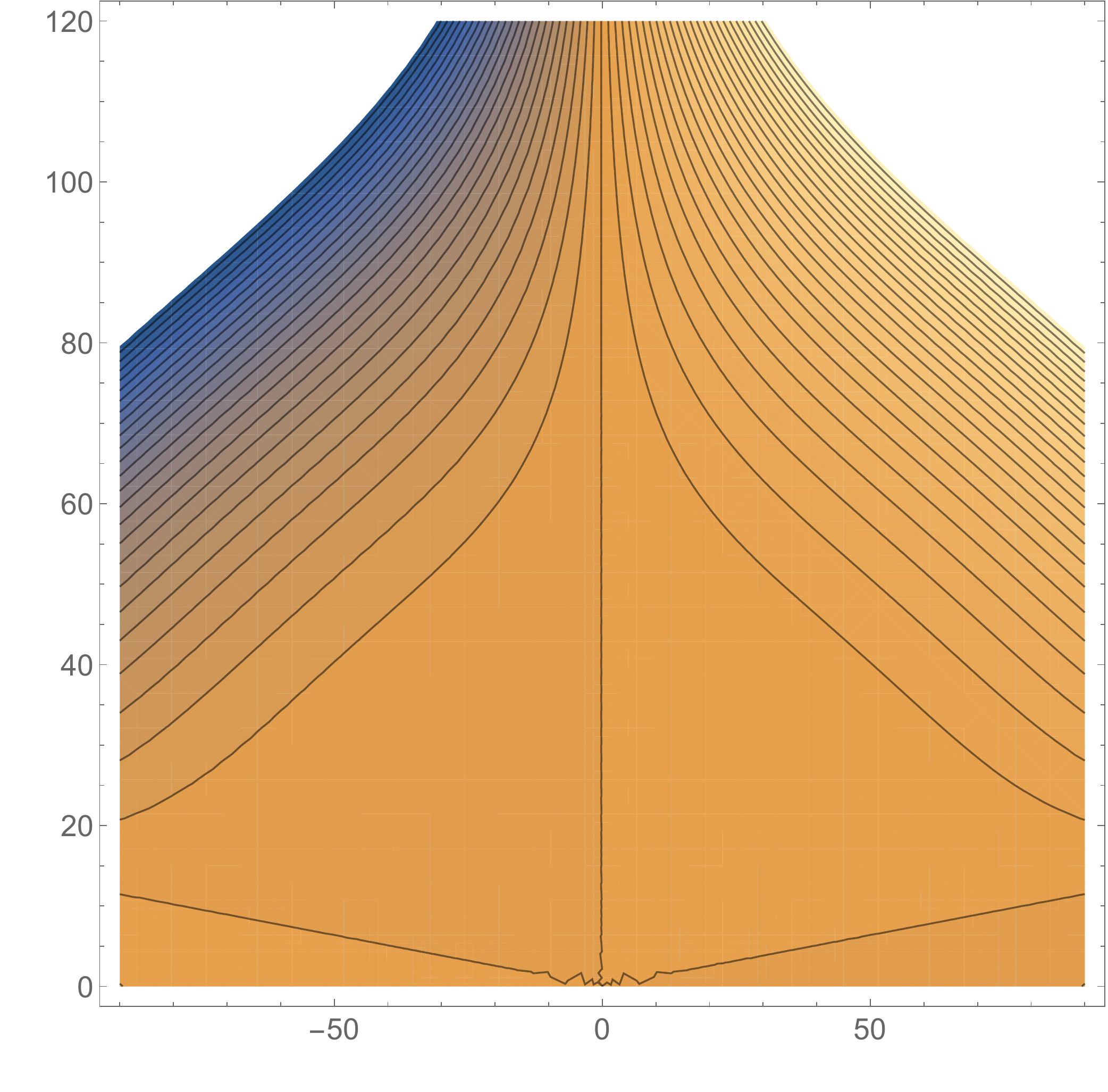}
  \endminipage\hfill
\minipage{0.32\textwidth}%
  \includegraphics[width=\linewidth]{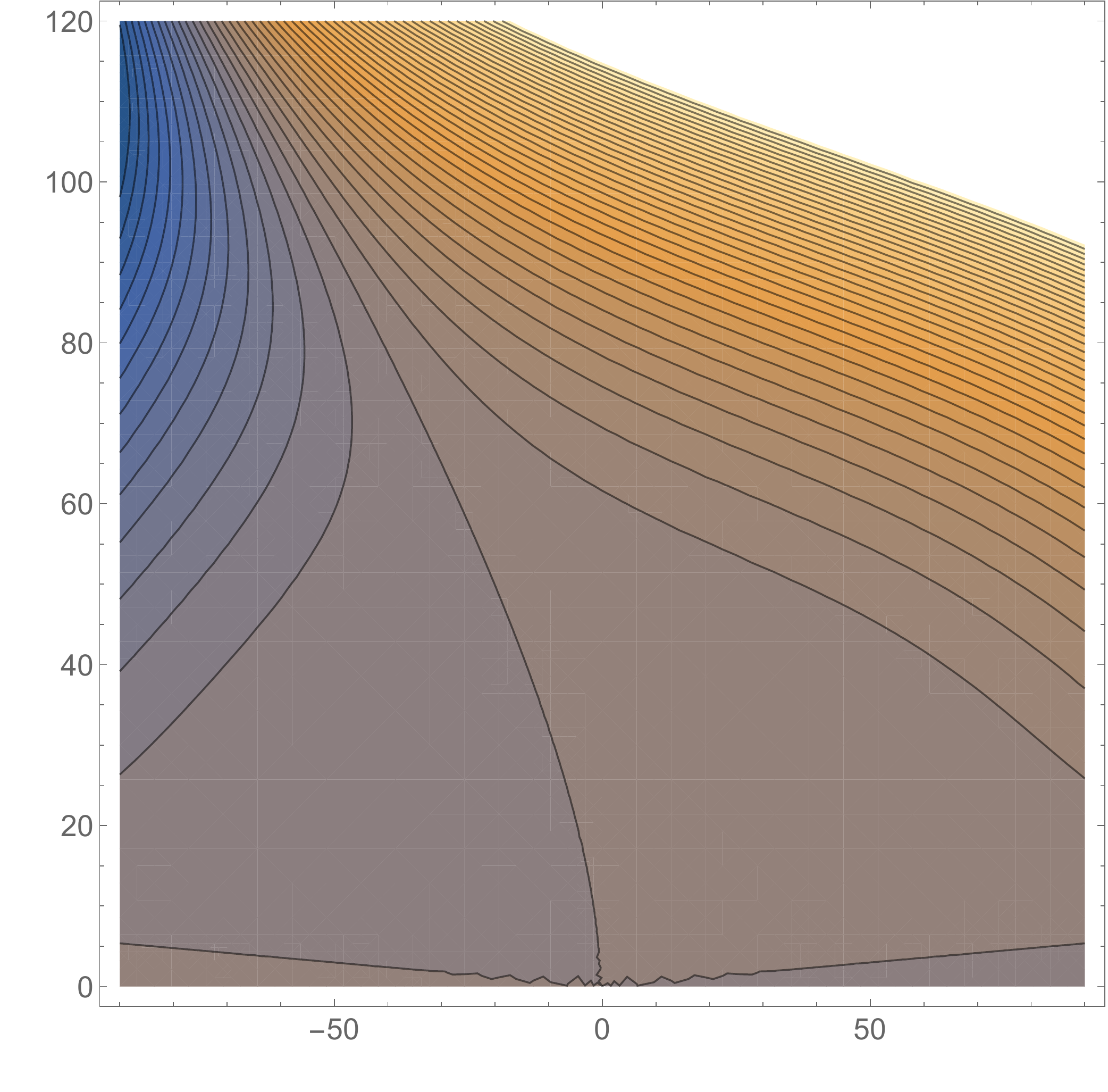}
 \endminipage
\caption{The figures are made respectively with $(\epsilon = 1,0,-1)$ with fixed parameter $v_0=0.02$ and unitary value for mass and charge with 50 contour lines in the ranges r[-90,90] and z[0, 120].}
\end{figure}
We notice that a small value induces to outgoing lines from the center as expected. In the center figure, we notice that for time-like curves, the magnetic lines distort the path of a test charged particle away from the center of the galaxy.

In the same sense, we investigate a Toomre-Kuzmin-like potential since we are dealing with an axisymmetric system, which has the form
$$U(r)= 1-\frac{\alpha}{\sqrt(r^2+z^2)}\;,$$
where $\alpha$ is a unitary free parameter to guarantee the correct units, and straightforwardly, one can get the expression
$$d\Phi_{eff}= \frac{2r}{\alpha^2}\{\epsilon+ \frac{2}{\alpha m^2 r^2}\left(z-\sqrt(r^2+z^2)\right)^2 + \frac{2\sqrt(2)q \alpha L}{m}\left(\frac{z-\sqrt(r^2+z^2)}{\sqrt(r^2+z^2)}\right)\} (r^2+z^2)\left(rdr+zdz\right) \;,$$
And, after a long algebra, we can write the form of the effective potential felt by charged particle with mass moving with total angular momentum $L$
$$\Phi_{eff}(r,z)= P(z)r^3 + Ur^5 - K(z) \ln |z| + V(r,z) \sqrt(r^2+z^2) + I(r,z) + C(z)\ln|r+\sqrt(r^2+z^2)|\;.$$
where we denote the following terms
\begin{eqnarray*}
  P(z)&=&\frac{2\epsilon}{3 \alpha^2} z^2 + \frac{4}{m^2\alpha^3} z^2 + \frac{3\epsilon}{2 \alpha^2} +\frac{3}{m^2 \alpha^3}- \frac{3\sqrt(2)}{m \alpha} q L - \frac{4\sqrt(2)}{3m\alpha} qL z^2 \\
  U &=& \frac{1}{5} \left(\frac{2 \epsilon}{\alpha^2} + \frac{4}{m^2\alpha^3} \right) - \frac{4}{5} \frac{\sqrt(2)}{m\alpha}qL\\
  K(z)&=&\frac{3}{m^2\alpha^3} z^5 + \frac{\sqrt(2)}{2 m \alpha} qL z^3  \\
  V(r,z)&=&\frac{4 r z}{m^2 \alpha^3} \left(r^2+z^2\right) + \frac{3}{m^2\alpha^3} z^3 + \frac{\sqrt(2)}{2} \frac{qL}{m\alpha} rz\left(2r^2+z^2\right),\\
  I(r,z)&=& \frac{8r}{\alpha^3 m^2} z^4 \\
  C(z)&=&\frac{3}{m^2\alpha^3} z^5 - \frac{\sqrt(2)}{2} \frac{qL}{m\alpha} z^3\\
  \end{eqnarray*}

\begin{figure}[!htb]
\minipage{0.32\textwidth}
  \includegraphics[width=\linewidth]{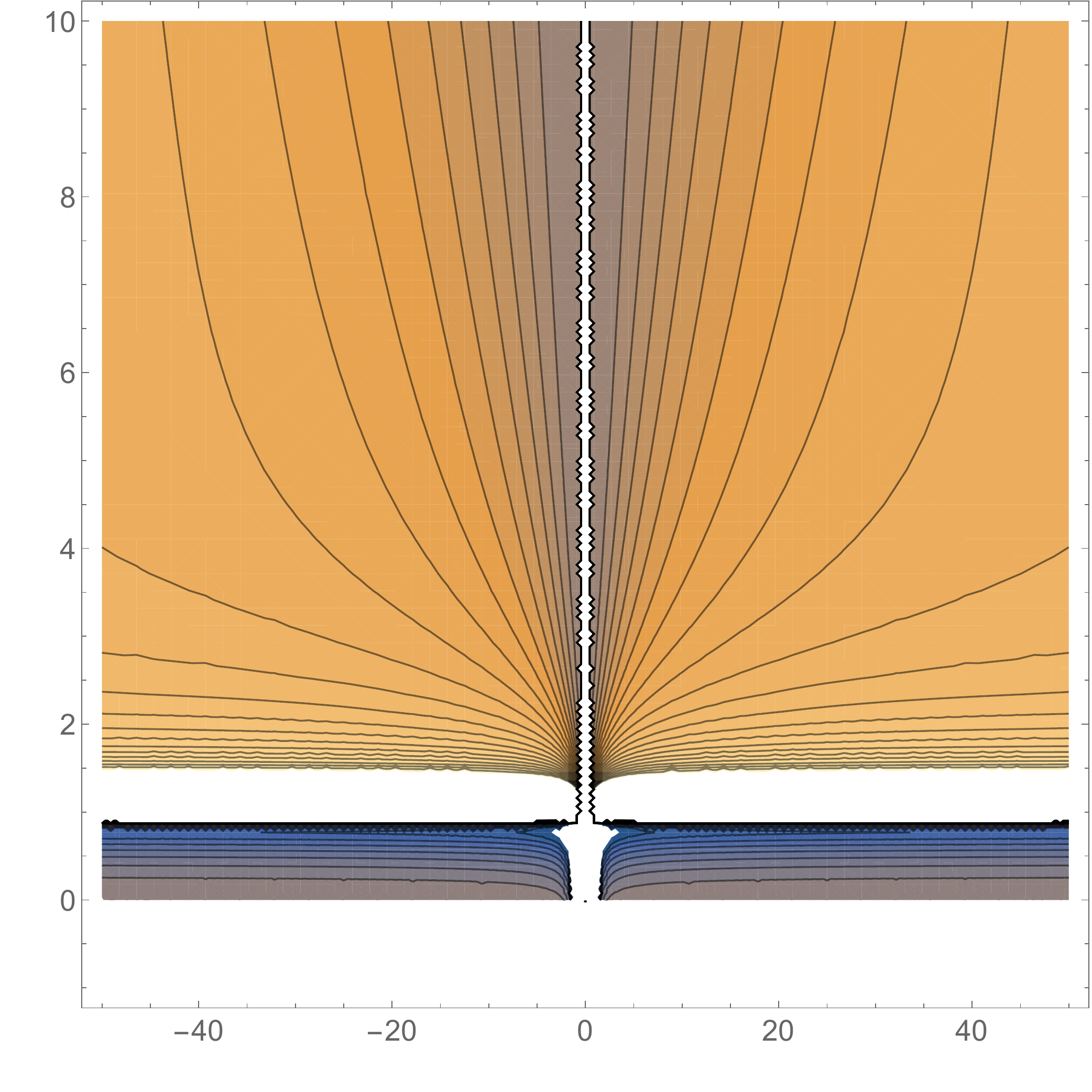}
  \endminipage\hfill
\minipage{0.32\textwidth}
  \includegraphics[width=\linewidth]{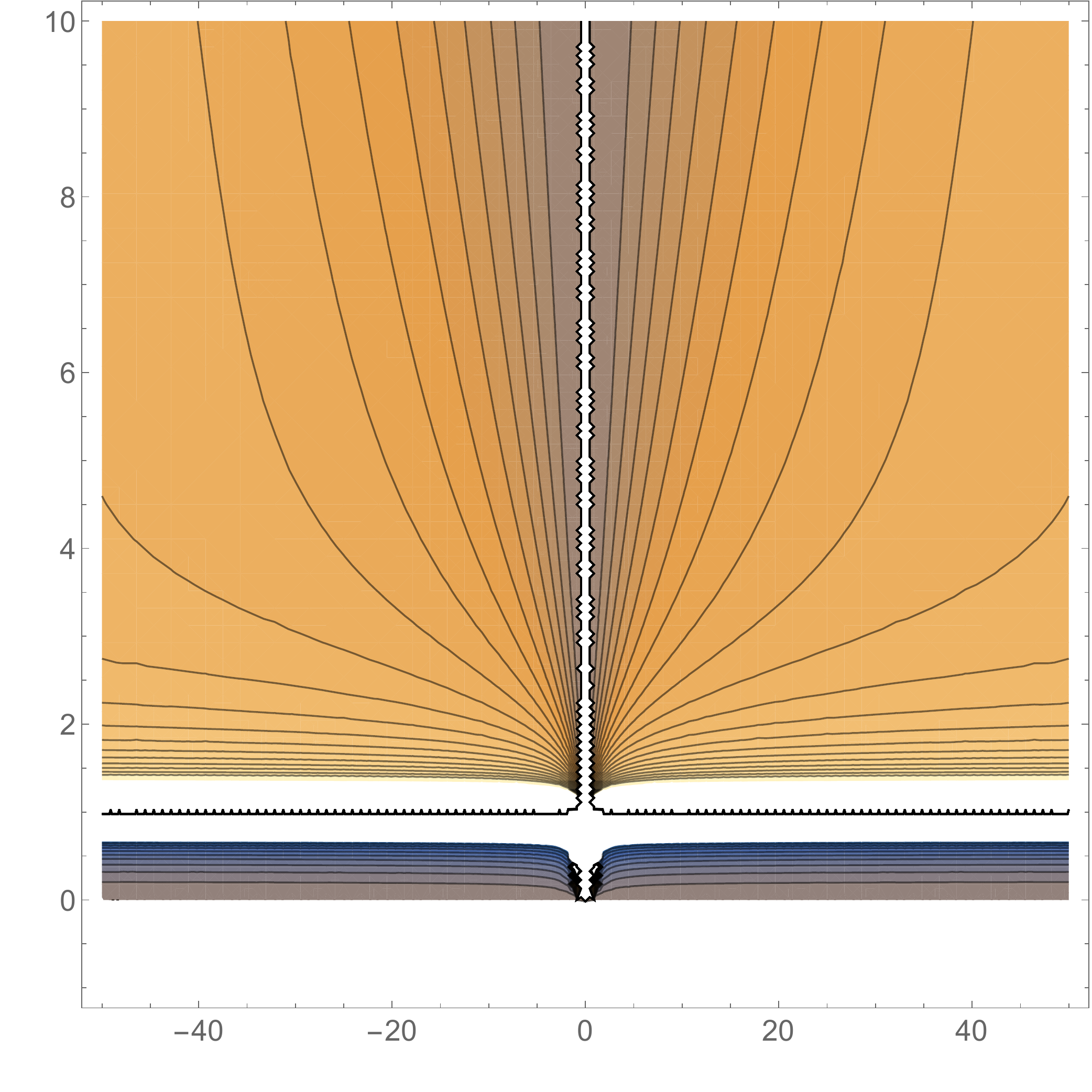}
  \endminipage\hfill
\minipage{0.32\textwidth}%
  \includegraphics[width=\linewidth]{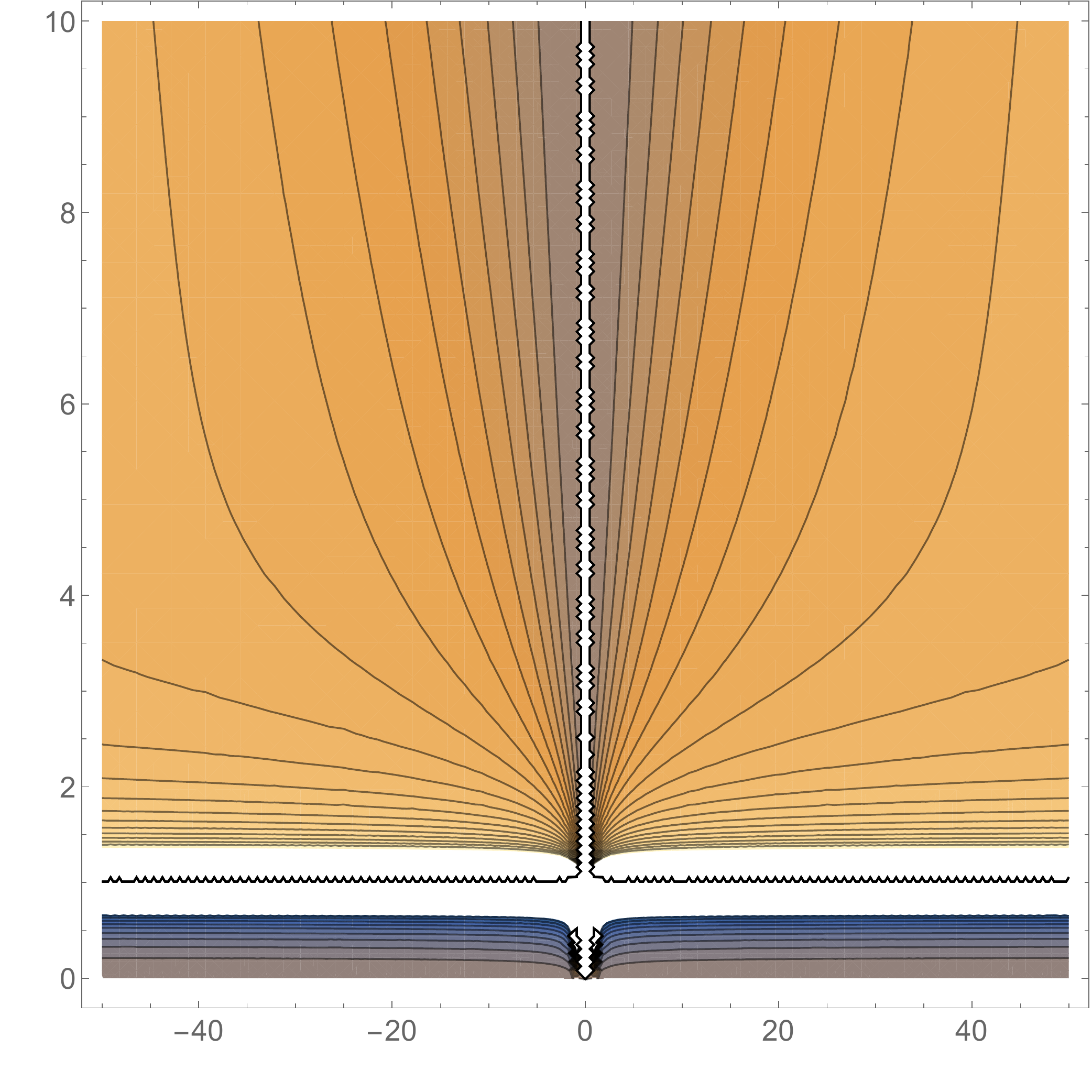}
 \endminipage
\caption{The figures are made respectively with $(\epsilon = 1,0,-1)$ with fixed parameters $m=1$, $q=1$, $L=1$, and $\alpha=1$, for 30 contour lines in ranges r[-50,50] and z[-1,10].}
\end{figure}
In this situation, we do not have any considerable difference between the figures and around the origin it is possible to check the singularity and the lines away to the center.

On the other hand, we can express the effective potential directly related to energy. In doing so, we use the  relations in Eqs.(\ref{eq:energy}),  (\ref{eq:angularmomentum})  and (\ref{eq:circuprecess}) that give three  linear differential equations,
involving  the four  unknowns $\dot{x}^{\alpha}$. It is  possible to  study the motion of  test  particles  with only  these  relations, if we limit ourselves to the  particular  case  of    equatorial trajectories, i. e.    $z = 0$. Indeed, since the gravitational configuration is symmetric with respect to the equatorial plane, a particle with initial state $z=0$ and $\dot z=0$ will remain confined to the equatorial plane which is, therefore, a geodesic plane. Substituting the  conserved quantities of Eqs.(\ref{eq:energy}) and (\ref{eq:angularmomentum}) into Eq.(\ref{eq:circuprecess}), we  find
\begin{equation}
  \dot{r}^2 + \Phi_{eff} = \frac{E^2}{m^2c^2}, \label{eq:motion}
\end{equation}
where
\begin{equation}
  \Phi_{eff}(r) \equiv \frac{L^2}{m^2r^2}\left(1 - \frac{qA_{\varphi}}{Lc}  \right)^2 e^{4\phi} + \Sigma e^{2\phi}
  \label{eq:effectivepot}
\end{equation}
is an effective  potential.  We assume the convention that the  positive  value  of  the  energy  corresponds to  the positivity of the solution $E_{\pm} = \pm m c\Phi_{eff}^{1/2}$.  Consequently, $E_{+} = - E_{-} =  m c\Phi_{eff}^{1/2}$.

The    motion  of  charged test particles  is  governed  by  the  behavior of  the  effective potential in Eq.(\ref{eq:effectivepot}). The  radius of  circular  orbits and  the corresponding values of  the energy $E$ and  angular momentum  $L$ are  given by  the  extrema of the  function $\Phi_{eff}$. Therefore, the  conditions  for  the  occurrence of  circular  orbits  are
\begin{equation}
  \frac{d\Phi_{eff}}{dr} =0, \quad  \Phi_{eff}= \frac{E^2}{m^2c^2}.
  \label{eq:circularconditios}
\end{equation}
Thus,  by  calculating the condition in Eq.(\ref{eq:circularconditios})  for  the  effective  potential in Eq.(\ref{eq:effectivepot}),  we  find the angular momentum  of the particle in   circular  motion
\begin{equation}
  L_{c \pm} =  \frac{q A_{\varphi}}{c}
                  +   \frac{q r A_{\varphi,r} e^{\phi} \pm \sqrt{ \left(q r A_{\varphi,r} e^{\phi}\right)^2
                  -   4 \Sigma c^2m^2r^3 \phi_{,r} \left(2 r \phi_{,r} - 1\right)} }
                     {2 c  e^{\phi} \left(2 r \phi_{,r} - 1\right)}.   \label{eq:circangularmomentum}
\end{equation}
Conventionally,  we  can associate  the plus and minus signs in the subscript of the notation $L_{c\pm}$ to  dextrorotation and  levorotation, respectively. Moreover, by  inserting the  value  of  the  angular momentum
in Eq.(\ref{eq:circangularmomentum}) into the  second  equation of Eq.(\ref{eq:circularconditios}), we  obtain the  energy  $E_{\quad c  \pm}^{(\pm)}$ of  the particle  in a  circular  orbit as
\begin{equation}
   E_{\quad c  \pm}^{(\pm)} = \pm\, m c e^{\phi} \left(\Sigma + \xi^{(\pm)}_{\quad c}\right)^{1/2}
   \label{eq:energycir},
\end{equation}
where
\begin{equation}
    \xi^{(\pm)}_{\quad c}= \frac{\left[ q r A_{\varphi,r} e^{\phi} \pm \sqrt{ \left(q r A_{\varphi,r}
    e^{\phi}\right)^2 - 4 \Sigma c^2 m^2 r^3   \phi_{,r} \left(2 r \phi_{,r} - 1\right)} \right]^2 }
                            {4 m^2 c^2 r^2   \left(2 r \phi_{,r} - 1\right)^2}.
                            \label{eq:xi-general}
\end{equation}
Therefore,  each  sign of the value of  the  energy corresponds to two kind of  motions (dextrorotation and  levorotation) indicated  in  Eqs.(\ref{eq:energycir})  and (\ref{eq:xi-general}) by  the  superscripts $(\pm)$.

\section{Perihelion advance in a  conformastatic magnetized  spacetime}
One of the most important tests of general relativity and modified theories of gravitation in astrophysical scale is the perihelion advance of celestial objects. In this section, we present
 the analytic expressions which determine the perihelion advance of charged test particle, moving in a conformastatic spacetime under the presence of a magnetic field. Starting with the first integral in Eq.(\ref{eq:circuprecess}),   we
restrict the analysis to the  motion of a particle on the
plane with $z=0$. Then,  we  have \begin{equation}
   \Bigg(\frac{dr}{d\varphi}\Bigg)^2 = - r^2 \Bigg[ 1  +  \frac{m^2r^2}{\big(L - \frac{q}{c}A_{\varphi}\big)^2}
                                   \Bigg(\Sigma (1- U)^2  - \frac{E^2 }{m^2 c^2} (1-U)^4 \Bigg)
    \Bigg],
   \label{eq:orbit}
\end{equation}
where all the  quantities are evaluated at $z=0$ and we  have  used  the  expressions  for the
energy  and  angular momentum of  the  particle  given  by Eqs.(\ref{eq:energy}) and
(\ref{eq:angularmomentum}), respectively.  With  the  change of   variable $u=1/r$,  Eq.(\ref{eq:orbit}) can be transformed  into
\begin{equation}
   \frac{d^2u}{d\varphi^2}+  u^2 = F(u)
   \label{eq:orbitu}
\end{equation}
where
\begin{equation}
  F(u) \equiv \frac{1}{2} \frac{d G}{du},
  \label{eq:explicF}
   \end{equation}
    and
   \begin{equation}
 G(u) \equiv \frac {1}   {\left(1 -   \frac{q A_{\varphi}} {c L}  \right)^2}
                \left[  \frac{E^2}{c^2L^2} (1 - U )^4 - \frac{\Sigma m^2}{L^2}(1 -U)^2 \right] \ .
\label{eq:explicG}
\end{equation}
 Accordingly,  by  following  the   procedure proposed  in  \cite{Harko2010},  we  have for
the  resulting perihelion advance
\begin{equation}
   \delta \varphi = \pi \Bigg(\frac{dF}{du}\Bigg)_{u=u_0},
   \label{eq:advance}
   \end{equation}
  where  $u_0$ is  the radius of  a nearly  circular  orbit,  which is  given  by  the  roots  of
the  equation $F(u_0)=u_0$. In Eq.(\ref{eq:advance}), we have shown the procedure to obtain an expression for the  perihelion advance of a charged test  particle  in a generic  conformastatic spacetime with a magnetic field. We now illustrate the  results by  considering a  particular  conformastatic spacetime generated  from the  harmonic  potential of a punctual mass
\begin{equation}
    U(r,z)=-\frac{G M }{c^2 R},  \quad R^2= r^2 + z^2
    \label{eq:Upotential2}.
    \end{equation}
Thus,  by  inserting  Eq.(\ref{eq:Upotential2}) into  Eq.(\ref{eq:explicF})
we obtain  for $F(u)$
\begin{equation}
F(u)= \frac{\left[  \frac{2E^2 GM}{c^4 L^2}    \left( 1 + \frac{G M}{c^2} u \right)^3
                        - \frac{\Sigma m^2 G M }{c^2 L^2}  \left( 1 + \frac{G M}{c^2} u  \right)
                         \right]}{\left( 1  - \frac{q \sqrt{G} M}{c L}  \right)^2}.
 \label{eq:Fequation}
\end{equation}
Accordingly,  the  perihelion advance of  a particle  in this spacetime  is  given  by
\begin{equation}
\delta \varphi= \pi \frac{\left[  \frac{6 E^2 G^2 M^2}{c^6 L^2}  x_0^2
                        - \frac{\Sigma m^2 G^2 M^2}{c^4 L^2}
                         \right]}{\left( 1  - \frac{q \sqrt{G} M}{c L}  \right)^2},
                         \label{eq:advance1}
\end{equation}
where the term
\begin{equation}
x_0 \equiv  1 + \frac{G M}{c^2} u_0
\end{equation}
satisfies  the  equation
\begin{eqnarray}
2 E^2G^2 M^2 x_0^3
         - \left[\Sigma m^2 G^2 M^2 c^2
         + c^6L^2 \left(1 - \frac{q\sqrt{G} M}{cL}\right)^2\right] x_0 \label{eq:x03equation}\nonumber\\
         + c^6L^2 \left(1 - \frac{q \sqrt{G}M}{cL} \right)^2 = 0.
\end{eqnarray}
Thus, by  inserting  the real  solution of Eq.(\ref{eq:x03equation}) into Eq.(\ref{eq:Fequation}), we find that
the perihelion advance of the test particle orbit is given by
\begin{equation}\label{eq:periheliondrift}
\delta{\varphi} = \pi \frac{ \psi_0 - k_2^2}{ Q^2},
\end{equation}
where
\begin{equation*}
 \psi_0 \equiv \frac{\left[ 6\left(Q^2 + k_2^2\right)
                  + \left[ 54Q^2k_1 \left( -1 + \sqrt{1 -\frac{6\left(Q^2 +
k_2^2\right)^3}{81Q^4k_1^2}  }    \right) \right]^{2/3}      \right]^2}
{ 6\left[ 54Q^2k_1 \left( -1 + \sqrt{1 -\frac{6\left(Q^2 +
k_2^2\right)^3}{81Q^4k_1^2}  }    \right) \right]^{2/3}  },
\end{equation*}
with
\begin{eqnarray*}
 k_1^2=\frac{E^2G^2M^2}{c^6L^2},\quad
 k_2^2=\frac{\Sigma m^2 G^2M^2}{c^4L^2},
   \end{eqnarray*}
and also
\begin{equation*}
 Q^2= \left(1 - \frac{q\sqrt{G}M}{cL}\right)^2\;.
\end{equation*}
Notice  that when $q=0$ (and,  consequently $Q=1$) we  get  the  case in which  Eq.(\ref{eq:advance1}) describes  the  perihelion  advance of
a  neutral  particle. Actually, we restrict ourselves to the neutral case, since objects like planets, asteroids and comets are neutral on average and the consideration of a charge is hardly significant. Moreover, this neutrality is essentially due to the influence of the solar wind, but a global net charge, e.g. in stars, is still on discussion \cite{neslusan}.

\begin{table*}
  \centering
  \begin{minipage}{140mm}
  \caption{Comparison between the values for secular precession of inner planets in units of arcsec/century $(''.cy^{-1})$ of the standard (Einstein) perihelion precession
	$\delta \varphi_{eins}$ \citep{Wilhelm} for
neutral test particles (planets, asteroids/comets)  in  the conformastatic magnetized  spacetime of a punctual mass $\delta \phi_{eff}$.
The data for $\delta \varphi_{obs}$ stands for the secular observed perihelion precession in
units of arcsec/century adapted from \citep{nambuya} by adding a supplementary precession correction from EPM2011 \citep{Pitjeva,Pitjev}.
In addition, the results for the NEOS 433 Eros, 3200 Phaethon  and 2p/Encke comet are also presented. The mass of the 2p/Encle comet as $m=3.85\times10^{13} kg$ was estimated with a bulk density $\rho=0.5 g.cm^3$  as shown in \citep{Fernandez}. }
\label{table1}
\begin{tabular}{@{}llrrrrr@{}}
  \hline
   Object             &    $\delta \varphi_{obs}$ \;\;\;\;\;\;\;\;\;\;\; & \;\;\;\;\;\;\;\;$\delta \varphi_{eins}$ \;\;\;\;\;\;\; &
   \multicolumn{4}{c}{\;\;\;\;\;\;\; $ \delta\varphi_{eff}$ \;\;\;\;\;\;\;\;\;\;\;\;\;\;\;\;\;\;\;\; $\beta_0$}   \\

  \hline
  Mercury            & 43.098 $\pm$ 0.503        &   42.97817       & \;\;\;\;\;\;\; 42.9782  &\;\;\;\;\;0.7605$\times\; 10^{-4}$ \\
                     &                           &                  &&  \\

  Venus              & 8.026  $\pm$ 5.016        &   8.62409        &\;\;\;\;\;\;\;  8.62425   & \;\;\;\;\;0.1426$\times\; 10^{-2}$ \\
                     &                           &                  & &  \\

  Earth              & 5.00019  $\pm$ 1.00038    &   3.83848        & \;\;\;\;\;\;\;3.83944    & \;\;\;\;\;0.4375$\times\; 10^{-2}$ \\
                     &                           &                  & &  \\

  Mars               & 1.36238  $\pm$ 0.000537   &   1.35086        & \;\;\;\;\;\;\;1.36980    & \;\;\;\;\;0.3729$\times\; 10^{-1}$\\
                     &                           &                  & &  \\

  433 Eros              & 1.60    &   1.57317                      & \;\;\;\;\;\;\;1.58668      & \;\;\;\;\;0.2906$\times\; 10^{-1}$ \\
                        &                       &                  & & \\

  3200 Phaethon         & 10.1    &   10.1201                      &\;\;\;\;\;\;\;10.1213     &\;\;\;\;\; 0.3499$\times\; 10^{-2}$ \\
                        &                       &                  & & \\

  2p/Encke              & 1.9079    &   1.868                      &\;\;\;\;\;\;\;1.92833      & \;\;\;\;\;0.5623$\times\; 10^{-1}$ \\
                        &                       &                  & & \\
  \hline
\end{tabular}
\end{minipage}
\end{table*}

In order to get a real use of Eq.(\ref{eq:periheliondrift}), we follow the procedure presented in \cite{Harko2010}. First,
we  rewrite both the angular momentum (\ref{eq:circangularmomentum}) and the energy
(\ref{eq:energycir}), which depend on the radial distance $r$, in terms of the parameters that describe the orbit of  rotating test particles.
For the radial distance, one can use the ellipse formula in the Euclidean plane as
\begin{equation}\label{eq:radiuseuclidean}
    r= \frac{ s (1-\epsilon^2)}{1+\epsilon \cos\varphi}\;,
\end{equation}
where $s$ is the semimajor axis and  $\epsilon$ the eccentricity of the orbit.
Moreover, we can rewrite Eq.(\ref{eq:periheliondrift}) by using physical units related to observations as
\begin{equation}\label{eq:periheliondrift2}
\delta{\varphi^{\star}} = \pi \gamma^{\star}\frac{ \left(\psi_0 - k_2^2\right) s^2}{ Q^2 M_{\odot}T^2},
\end{equation}
where we have introduced the solar mass $M_{\odot}$ and the period $T$ of the rotating body.
The parameter $\gamma^{\star}= \frac{180/\pi}{3600}T$ allows us to transform units from radians to
(secular) degrees. Moreover, in order to obtain a real effective advance $\delta \varphi_{eff}$ and to alleviate the error propagation, we define a deviation formulae  away from general relativity standard result $\delta \varphi_{eins}$ induced by the coupled Einstein-Maxwell fields as
\begin{equation}\label{eq:periheliondrift3}
\delta{\varphi_{eff}} = \delta \varphi_{eins} \pm \beta_0 \delta{\varphi^{\star}}\;,
\end{equation}
where a dimensionless parameter $\beta_0$ measures the
tiny variation of the orbits through time. As we have checked in the studied cases in table (\ref{table1}), a variation of $\beta_0$ must not exceed $10\%$ of the ratio between the Einstein-Maxwell contribution $\delta\varphi^{\star}$ and observations $\delta\varphi_{obs}$.

When applied to the observational data \cite{nambuya} plus a supplementary precession corrections from EPM2011 \cite{Pitjeva,Pitjev}, one can test
Eq.(\ref{eq:periheliondrift3}). Thus, we  obtain the results presented in table (\ref{table1})  for the perihelion precession of inner planets of the Solar system, two NEO's asteroids named 433 Eros and 3200 Phaethon, and NEO 2p/Encke comet. The data for the astrophysics parameters of planets like semimajor axis, eccentricity, period and mass, can be found  in JPL solar system dynamics (http://ssd.jpl.nasa.gov/?planets) and for asteroids and comets, in JPL small body database (http://ssd.jpl.nasa.gov/sbdb.cgi). The orbital periods are in units of years.

As shown in table (\ref{table1}), the theoretical results match the observations, and a slight improvement is obtained as compared to the standard Einstein gravity which turns our model closer to the observations. We conclude that the gravitational interaction generated by the magnetic field of the central body can play an important role in astrophysical observations. It is worth to say that the values of $\alpha$ seem to be sensitive to the variation of the eccentricity of the orbits and the mass of the object as seen in the studied cases and the values have a close resemblance to PPN parameters that have a bound $|2\gamma - \beta -1| < 3 \times 10^{-3}$ \cite{Will}.

In addition, some other considerations must be noted. The constant $\Sigma$ enters explicitly the expression for the perihelion advance in Eq.(\ref{eq:circuprecess}),  and it represents null, time-like and space-like curves. For $\Sigma=0$, we do not have a  solution since Eq.(\ref{eq:periheliondrift3}) diverges. For time-like trajectories, $\Sigma=-c^2$ no physical results are obtained, because in the corresponding Newtonian limit a differential equation is obtained, whose solution implies that $r$ is negative. Moreover, no significant differences were found for different values of the charge of the order  $q/m \sim 10^{-3}$, which is the value where the behavior of the energy and angular momentum becomes affected by the presence of the effective charge. In the same sense, no differences could be found when using both solutions for the angular momentum $L_{c\;\pm}$ and energy $E_{c\;\pm}$.

\section*{Conclusion}

In  this  work we  have  shortly shown   the  characteristics of  the  motion of  a  charged particle  along  circular orbits in a  spacetime described  by   a  conformastatic solution  of the Einstein-Maxwell  equations.  As  a  particular  example  we  have  considered the  case  of  a  charged particle moving  in the gravitational  field   of  a  punctual  source placed at the origin of coordinates. Our  analysis  is  based  on the  study of the behavior of an  effective  potential that  determines  the  position and  stability  properties  of  circular  orbits. We also have investigated the behaviour of effective potentials.

In addition,  we  have  also  calculated  an  expression  for    the  perihelion advance  of a test  particle in a general   magnetized conformastatic  spacetime obtaining a good agreement with the observed values for the perihelion of inner Solar planets and some selected NEO asteroids. It is worth noting that all results presented were obtained with the initial assumption of a neutral particle, in accordance with the fact that planets are largely neutral. Specifically, in the perihelion drift, we find that the differences between a neutral particle and a charged particle are sightly small, when realistic values for the effective charge are used. This means that the electromagnetic interaction between the charge and the central magnetized body does not seriously affect the value of the perihelion advance. Nevertheless, the magnetic field enters explicitly the metric components and, consequently, affects the motion of neutral test particles through the gravitational interaction. This explains why the numerical predictions of the perihelion advance generated by  a punctual magnetized mass are in better agreement with observations than the predictions of Einstein's theory alone. As a future prospect, we will apply the Poincar\'{e} surface-of-section method for analyzing weakly perturbed Hamiltonian conformastatic systems.

       \end{document}